\documentclass[showpacs,twocolumn,prb]{revtex4}
\usepackage{amsfonts}
\usepackage[latin9]{inputenc}
\usepackage{amsmath}
\usepackage{amssymb}
\usepackage{hyperref}
\usepackage{graphicx}

\begin{document}

\title{Multiband tunneling in trilayer graphene}
\date{\today }
\author{B. Van Duppen}
\email{ben.vanduppen@ua.ac.be}
\affiliation{Department of Physics, University of Antwerp, Groenenborgerlaan 171, B-2020 Antwerp, Belgium}
\author{S. H. R. Sena}
\affiliation{Department of Physics, University of Antwerp, Groenenborgerlaan 171, B-2020 Antwerp, Belgium}
\affiliation{Departamento de F\'{i}sica, Universidade Federal do Cear\'{a}, Fortaleza, Cear\'{a}, 60455-760, Brazil}
\author{F. M. Peeters}
\affiliation{Department of Physics, University of Antwerp, Groenenborgerlaan 171, B-2020 Antwerp, Belgium}
\affiliation{Departamento de F\'{i}sica, Universidade Federal do Cear\'{a}, Fortaleza, Cear\'{a}, 60455-760, Brazil}

\begin{abstract}
The electronic tunneling properties of the two stable forms of trilayer graphene (TLG), rhombohedral ABC and Bernal ABA, are examined for $pn$ and $pnp$ junctions as realized by using a single gate (SG) or a double gate (DG). For the rhombohedral form, due to the chirality of the electrons, the Klein paradox is found at normal incidence for SG devices while at high energy interband scattering between additional propagation modes can occur. The electrons in Bernal ABA TLG can have a monolayer- or bilayer-like character when incident on a SG device. Using a DG however both propagation modes will couple by breaking the mirror symmetry of the system which induces intermode scattering and resonances that depend on the width of the DG $pnp$ junction. For ABC TLG the DG opens up a band gap which suppresses Klein tunneling. The DG induces also an unexpected asymmetry in the tunneling angle for single valley electrons.
\end{abstract}

\maketitle

\section{Introduction}

The discovery of a one atom thick layer of carbon atoms, graphene, opened up an entire new field in the condensed matter world\cite{Novoselov2004}. The electronic properties of graphene have an intriguing analogy with ultrarelativistic particles such as the linear electronic spectrum and the occurrence of Klein tunneling \cite{Katsnelson2006} which both have been experimentally verified \cite{Sprinkle2009,Stander2009,Young2009}. This analogy leads to the introduction of the concept of pseudospin \cite{Katsnelson2006} giving the carriers a chirality which is closely linked to these phenomena.

Graphene multilayers though being bound by a weak Van der Waals force posses energy spectra that are fundamentally different from the monolayer case \cite{Partoens2007,Min2008}. The low energy spectrum is no longer linear and different modes of propagation become possible due to the presence of multiple energy bands that appear as a consequence of the increasing number of atoms in the unit cell. It was shown that the behavior of these bands \cite{Guinea2006,Aoki2007}, their response to an external applied gate voltage \cite{Avetisyan2010} as well as the transport properties of the system \cite{Jhang2011,Bao2011} strongly depend on the way the graphene sheets are stacked.

The transport properties of bilayer graphene (BLG), the thinnest multilayer structure, has been extensively studied \cite{Katsnelson2006,Snyman2007,Barbier2010}. In contrast to monolayer graphene (MLG), it was shown that, within the two band approximation, Klein tunneling does not occur, i.e. the transmission for normal incidence is practically zero because, due to pseudospinorial arguments, the negative energy states are cloaked from the positive ones\cite{Gu2011}. However, when the full-band Hamiltonian model is considered, it was recently reported \cite{VanDuppen} that normal transmission becomes possible for high enough potential barriers, which makes the higher energy bands available for conduction.

\begin{figure}[tb]
\centering
\includegraphics[width= 8.5cm]{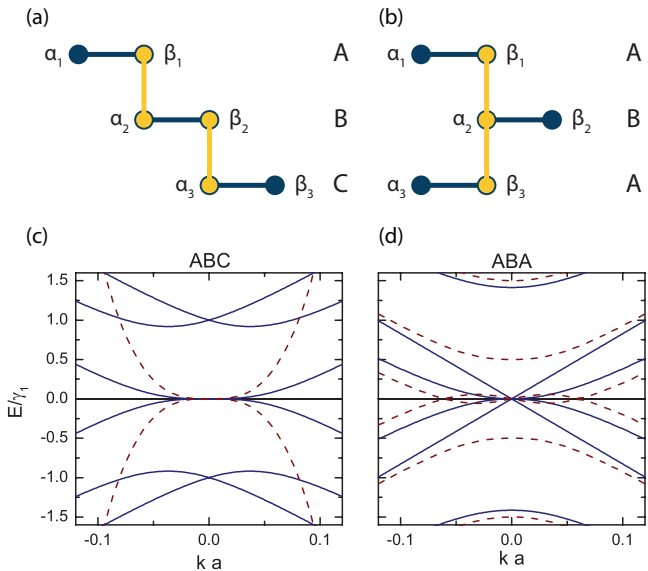}
\caption{(Colour online) (Top) Two different crystallographic structures showing the relative position of the sublattices $\alpha_i$ and $\beta_i$ for (a) ABC TLG and (b) ABA TLG. The interlayer hopping that is considered in this study is indicated by the yellow lines between the yellow marked atoms. (Bottom) Energy spectrum of (c) ABC TLG and (d) ABA TLG shown by the blue solid curves. The red dashed curves in (c) correspond to the spectrum of the two band Hamiltonian. The red dashed curves in (d) correspond to the spectrum of ABA TLG with interlayer bias $\protect\delta=0.5 \protect\gamma1 $.}
\label{CrisStructureAndBands}
\end{figure}

Recently, there has been a growing interest in the study of the electronic properties of trilayer graphene (TLG), since it constitutes the simplest multilayer system where both types of stacking order, Bernal (ABA) and rhombohedral (ABC) are possible. These stacking types are schematically shown in Figs. \ref{CrisStructureAndBands}(a) and (b), respectively. The mirror symmetric Bernal stacking is the most common and can be exfoliated from natural graphite since it shares its crystalline structure\cite{Bernal1924}. Although the ABC stacking is less common, it has recently been reported that 16$\%$ of the synthesized graphite \cite{Aoki2007} and around 15$\%$ of exfoliated TLG \cite{Jhang2011,Lui2011} has rhombohedral stacking. The way in which the layers are stacked influences strongly the energy spectrum of the system. It was demonstrated that for trilayer systems the occurrence of Klein tunneling depends on the staking order, being present only in ABC stacked trilayers \cite{Kumar2012, VanDuppen2012, VanDuppen2013}. Recent experiments have investigated the electronic transport in trilayers\cite{Zou2013, Campos}.

The tunneling problem of charge carriers in TLG is technically more complex which is the reason why only very recently this problem was tackled. Kumar \textit{et al.}\cite{Kumar2012} calculated the low energy tunneling through $pnp$ junctions. Unfortunately, the numerical results were shown not to be correct\cite{VanDuppen2012}. Here we extend this work to the more difficult regime of high energy and high potential when several propagation modes are present and thus multiple tunneling and reflection channels have to be taken into account. We consider both ABC and ABA TLG that are affected by a single gate (SG) that is able to locally vary the potential on each layer with the same value. Additionally, we calculate the multiband tunneling in the presence of a nanostructured double gate (DG) that affects the potential on each layer separately. We will show that this introduces interband scattering even at low energy in ABA TLG.

The paper is organized as follows. In Sec. \ref{Sec:Model} we introduce the full band continuum model used to describe ABC and ABA TLG and define the different potential profiles that are used in Sec. \ref{Sec:Potential}. In Sec. \ref{Sec:Eigenstates} we present in detail the formalism used to calculate the transmission and reflection probabilities for these multichannel systems. In Sec. \ref{Sec:Results} we show the results for transmission, reflection and conductance for both stacking sequences and considered different potential profiles. Finally, in Sec. \ref{Sec:Conclusion} we summarize our main conclusions.

\section{The model}\label{Sec:Model}

\subsection{ABC Trilayer}

For ABC trilayer, the effective Hamiltonian near the Dirac point in one valley can be calculated using the tight binding formalism. Considering only nearest neighbor interlayer transitions, marked in yellow in Fig. \ref{CrisStructureAndBands}(a), one obtains
\cite{Zhang2010a}
\begin{equation}
H_{ABC}=\hbar v_{F}\left[
\begin{array}{ccc}
\vec{\sigma}\cdot\vec{k} & \tau & 0 \\
\tau ^{\dag } & \vec{\sigma}\cdot\vec{k} & \tau \\
0 & \tau ^{\dag } & \vec{\sigma}\cdot\vec{k}%
\end{array}%
\label{Eq:HamABC}
\right] ,
\end{equation}%
with $\vec{\sigma}=(\sigma_x,\sigma_y)$ a vector of Pauli matrices, $v_{F}$ the Fermi velocity in monolayer graphene\cite{Partoens2007} ($v_F \approx 1.01 \times 10^6 m/s$), $\vec{k}$ the wave vector and $\tau$ describes the interlayer coupling which is given by
\begin{equation}
\tau =\frac{1}{\hbar v_{F}}\left[
\begin{array}{cc}
0 & 0 \\
\gamma _{1} & 0
\end{array}
\right] ,
\end{equation}
where\cite{Partoens2007} $\gamma _{1}=377meV$ is the interlayer hopping parameter. This Hamiltonian is written in the basis of the atomic orbital eigenfunctions
\begin{equation}
\Psi =\left( \psi _{\alpha_{1}},\psi _{\beta_{1}},\psi _{\alpha_{2}},\psi _{\beta_{2}},\psi_{\alpha_{3}},\psi _{\beta_{3}}\right)^{T} ,  \label{basisOrbitals}
\end{equation}%
where the indices indicate the sublattice associated with the respective eigenfunction. The energy spectrum of this Hamiltonian is depicted in Fig. \ref{CrisStructureAndBands}(c) by solid curves. The spectrum consists of six energy bands of which two touch each other at $k=0$. The two touching bands can be described approximately by the corresponding $2\times 2$ Hamiltonian\cite{Min2008, Nakamura2008}
\begin{equation}
H^{\prime }_{ABC}=\frac{\left( \hbar v_{F}\right) ^{3}}{\gamma _{1}^{2}}%
\left[
\begin{array}{cc}
0 & \left( k_{x}-ik_{y}\right) ^{3} \\
\left( k_{x}+ik_{y}\right) ^{3} & 0%
\end{array}%
\right] ,  \label{TwoBandTLG}
\end{equation}%
with dispersion relation
\begin{equation}
E _{l}=l \frac{\left( \hbar v_{F}\right) ^{3}}{\gamma _{1}^{2}}%
\left( \sqrt{k_{x}^{2}+k_{y}^{2}}\right) ^{3} ,
\end{equation}
where $l=\pm 1$. The energy spectrum of this Hamiltonian is shown in Fig. \ref{CrisStructureAndBands}(c) by the dashed curve making clear that this approximation is only valid for very small energy (i.e. $E/\gamma_1 < 0.2$) and near the Dirac point. The use of the approximate two band Hamiltonian allows for the extension of concepts as pseudospin  to the trilayer system, which was originally defined for monolayer graphene (MLG) \cite{Katsnelson2006}. This extension makes it possible to derive several electronic properties from the conservation of pseudospin analogous to monolayer graphene\cite{VanDuppen2013}.

\subsection{ABA Trilayer}\label{ABASection}

Following a similar approach as before, the effective Hamiltonian obtained by a tight-binding model considering only nearest neighbor interaction of the ABA trilayer is \cite{Partoens2007}:
\begin{equation}
H_{ABA}+\Delta =\hbar v_{F}\left[
\begin{array}{ccc}
\vec{\sigma}\cdot \vec{k}+\delta ^{\prime }I_{2} & \tau  & 0 \\
\tau ^{\dag } & \vec{\sigma}\cdot \vec{k} & \tau ^{\dag } \\
0 & \tau  & \vec{\sigma}\cdot \vec{k}-\delta ^{\prime }I_{2}%
\end{array}%
\right] ,  \label{Eq:HamABA}
\end{equation}%
in the same basis of orbital eigenfunctions as defined in Eq. $\left( \ref{basisOrbitals}\right) $. $I_{2}$ is the $2\times 2$ unit matrix and the $\delta ^{\prime }=\delta /\hbar v_{F}$ term corresponds to an externally induced interlayer potential difference of $\delta $ which is described by the term $\Delta$ at the left hand side of the equation. This is a $6\times6$ diagonal matrix given by
\begin{equation}
\Delta =Diag\left[ \delta ,\delta ,0,0,-\delta ,-\delta \right] \text{.}
\label{DeltaHamTerm}
\end{equation}
Despite the strong resemblance with the ABC Hamiltonian the ABA system is mirror symmetric with respect to the central layer. Therefore, a unitary transformation that combines the orbital eigenfunctions symmetrically and antisymmetrically transforms the Hamiltonian into a block diagonal form:\cite{Koshino2009b}
\begin{equation}
H_{ABA }^{\prime }+\Delta=\hbar v_{F}\left[
\begin{array}{ccc}
\vec{\sigma}\cdot \vec{k} & \delta ^{\prime }I_{2} & 0 \\
\delta ^{\prime }I_{2} & \vec{\sigma}\cdot \vec{k} & \sqrt{2}\tau  \\
0 & \sqrt{2}\tau ^{\dag } & \vec{\sigma}\cdot \vec{k}%
\end{array}%
\right] .  \label{HamABATransfro}
\end{equation}%
This new form of the Hamiltonian consists of a $2\times 2$ monolayer-like (top, left) and a $4\times 4$ bilayer-like (bottom, right) block that are connected by the parts responsible for the interlayer potential difference. When $\delta =0$\ the two blocks result in a superimposed linear (from the monolayer) and a hyperbolic (from the bilayer part) spectrum near the Dirac point as shown by the solid curves in Fig. \ref{CrisStructureAndBands}(d). In that case, electrons propagating in ABA TLG can propagate through two different modes, one monolayer-like and one bilayer-like mode. As long as the mirror symmetry remains intact, both modes will not interact and scattering between them is prohibited. When mirror symmetry is broken, e.g. by applying a different potential to every layer described by the term $\Delta$, interband scattering is possible. In Fig. \ref{CrisStructureAndBands}(d) the spectrum with non zero $\delta$ is shown by dashed curves. Due to symmetry breaking, the band crossing near the Dirac point is lifted making the linear monolayer-like bands become hyperbolic and the bilayer-like bands form a Dirac cone at the Dirac point while again crossing at higher wave vector.

\section{Electrostatic potential}\label{Sec:Potential}
In this paper we have considered two kinds of gates that influence the local electrostatic potential experienced by the electrons. The first one consists of a single gated device (SG) that causes a potential shift $V_0$ equal for all three layers. The second one is a double gated device (DG) that influences each layer differently inducing an interlayer potential difference $\delta$  between neighboring layers. Both systems are translational invariant in the $y$ direction.

\begin{figure}[tb]
\centering
\includegraphics[width = 8cm]{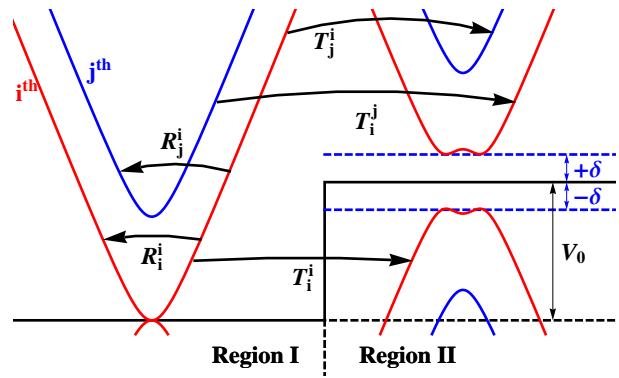}
\caption{(Colour online) Illustration of some of the different transmission and
reflection channels for a generic two band system with an $i^{th}$ and a $j^{th}$ band. The potential and interlayer bias in region II are $V_0$ and $\delta$ respectively, while both are zero in region I.}
\label{Fig:GenericTransAndRefl}
\end{figure}

The SG and DG act as a boundary for which we calculate the transmission and reflection probabilities defined in the previous section as function of the angle of incidence and the Fermi energy of the incident electron for different configurations of the devices. If the transmitted electrons are measured inside the gated region, it can be modeled as a single boundary corresponding to a $pn$ junction. The 1D potential profile is for convenience modeled by a step function
\begin{equation}
V_{pn}(x)=\left\{
\begin{array}{ccc}
0 & \text{if} & x<0 \\
V_{0}I_6+\Delta  & \text{if} & 0<x%
\end{array}%
\begin{array}{l}
\text{Region I} \\
\text{Region II}%
\end{array}%
\right. ,
\label{Eq:PotStep}
\end{equation}%
where $V_{0}$ is the height of the potential and corresponds to the SG term and the term $\Delta$ in this potential describes the effect of a DG by inducing a potential difference between the layers. This term is defined in Eq. $\left( \ref{DeltaHamTerm}\right)$, so the potential is $V_{0}$ for the middle layer and $V_{0}\pm \delta $ for the top and bottom layer. The $pn$ junction described here corresponds to the depicted schematic profile in Fig. \ref{Fig:GenericTransAndRefl}. The single valley approximation used in this paper assumes that the potential varies over a length scale larger than the in-plane interatomic distance\cite{Partoens2007} $a=0.142 nm$, but smaller than the electron wavelength.

When the electrons are measured outside the gated region, one can describe it as a $pnp$ junction or a potential barrier by
\begin{equation}
V_{pnp}(x)=\left\{
\begin{array}{ccc}
0 & \text{if} & x<0 \\
V_{0}I_6+\Delta  & \text{if} & 0\leq x\leq d \\
0 & \text{if} & x>d%
\end{array}%
\begin{array}{l}
\text{Region I} \\
\text{Region II} \\
\text{Region III}%
\end{array}%
\right. ,
\label{Eq:PotBar}
\end{equation}
where $d$ is the width of the gated region.

\section{Eigenstates, current density and transmission probability}\label{Sec:Eigenstates}

The eigenstates of these $6\times 6$ Hamiltonians are six-component spinors consisting of a superposition of three times two oppositely propagating or evanescent waves characterized by three distinct wave vectors which we call $k_{1}$, $k_{2}$ and $k_{3}$. For a system that is translational invariant in the $y$ direction, the energy and $k_{y}$ dependence of these wave vectors $k_{i}$ can be found from
\begin{equation}
\det \left[ H\left( k_{x},k_{y}\right) -EI_{6}\right] =0,  \label{DetEq}
\end{equation}%
which leads to a sixth power polynomial in $k_x$. The solution of it corresponds to the inversion of the energy spectrum. Since they solve the Dirac equation $H\Psi =E\Psi $, the eigenstate spinors can be written as a product of matrices
\begin{equation}
\Psi \left( x,y\right) =\mathcal{PE}\left( x,y\right) \mathcal{C},
\label{PECUitdrukking}
\end{equation}%
where $\mathcal{P}$ is a $6\times 6$ matrix expressing the relative importance of the different components of the spinor that can be constructed by solving the Dirac equation and
\begin{equation}
\mathcal{E}=Diag\left[
e^{ik_{1}x},e^{-ik_{1}x},e^{ik_{2}x},e^{-ik_{2}x},e^{ik_{3}x},e^{-ik_{3}x}%
\right] e^{-ik_{y}y},
\end{equation}%
where due to the translational symmetry in the $y$ direction, the $y$ dependency is incorporated in an exponential phase factor and will be ignored from this point on. We denote the six component vector $\mathcal{C}$ as
\begin{equation}
\mathcal{C}=\left[
a_{1}^{+},a_{1}^{-},a_{2}^{+},a_{2}^{-},a_{3}^{+},a_{3}^{-}\right] ^{T},
\end{equation}%
where the subscript $i$ refers to the corresponding wave vector and the superscript $+/-$ indicates the right/left propagating or evanescent states. The boundary conditions of the system under consideration will determine which of the components of the vector $\mathcal{C}$ are zero.

Using the continuity equation, one can derive the current density for a general graphene multilayer Hamiltonian with only nearest neighbor interlayer interactions and express it as a product of the above defined matrices. Such an Hamiltonian can be written in position representation as
\begin{equation}
H_{n}=-i\hbar v_{F}\vec{\alpha}.\vec{\nabla}+ \Gamma ,
\end{equation}%
where $\alpha _{x(y)}$ is a block diagonal matrix with $n$ Pauli matrices $\sigma_{x(y)}$ on the diagonal and $\Gamma $ consists of the other elements of the Hamiltonian that interconnect different atomic orbitals or induce a layer specific potential. Since the probability density is given by $\rho =\Psi ^{\dag }\Psi $, one can use the time dependent Dirac equation $i\hbar \partial _{t}\Psi =H\Psi $ to obtain the current density:
\begin{eqnarray}
i\hbar \partial _{t}\rho &=&i\hbar \left[ \left( \partial _{t}\Psi ^{\dag
}\right) \Psi +\Psi ^{\dag }\left( \partial _{t}\Psi \right) \right] \\
&=&i\hbar \left[
\begin{array}{c}
\left( -v_{F}\left( \vec{\nabla}\Psi ^{\dag }\right) .\vec{\alpha}-\Gamma
^{\dag }\Psi ^{\dag }\right) \Psi \\
+\Psi ^{\dag }\left( -v_{F}\vec{\alpha}.\vec{\nabla}\Psi +\Gamma \Psi \right)%
\end{array}%
\right] \\
&=&-i\hbar v_{F}\vec{\nabla}\left( \Psi ^{\dag }\vec{\alpha}\Psi \right) ,
\end{eqnarray}%
where use has been made of the hermiticity of $\Gamma $ and the $\vec{\alpha} $ matrices. The current density is hereby given by
\begin{equation}
\vec{j}=v_{F}\Psi ^{\dag }\vec{\alpha}\Psi .
\end{equation}%
Introducing the matrix notation of Eq. $\left( \ref{PECUitdrukking}\right) $, this expression becomes
\begin{equation}
\vec{j}=v_{F}\mathcal{C}^{\dag }\mathcal{E}^{\dag }\mathcal{P}^{\dag }\vec{%
\alpha}\mathcal{PEC}.
\end{equation}
Due to the properties of the spinor wave function, the matrix $\mathcal{\vec{A}}=\mathcal{P}^{\dag }\vec{\alpha}\mathcal{P}$ is diagonal consisting of traceless $2\times 2$ blocks that each correspond to a propagation mode. The resulting current density is
\begin{equation}
\vec{j}=v_{F}\sum_{j=1,\xi =\pm }^{3}\xi \left\vert a_{j}^{\xi }\right\vert
^{2}\mathcal{\vec{A}}_{j,j},  \label{CurrentDensityFormula}
\end{equation}%
where $\mathcal{\vec{A}}_{j,j}$ denotes the upper left element of the $%
j^{th} $ block.

In this paper we consider the transmission of electrons between regions of different electrostatic potential or interlayer potential difference. This introduces a spatially varying potential term in the Hamiltonian in Eq. $\left( \ref{DetEq}\right) $ that determines those regions. The boundary is parallel with the $y$ axis and we impose conservation of the transverse wave vector $k_{y}$. This reduces the system to a one dimensional problem with the conservation of Fermi energy $E$ and transverse momentum $p_y=\hbar k_{y}$, which defines the angle of incidence on the boundary. We are interested in the dependency of the transmission and reflection on these properties.

Suppose the electron is incident at the left side of the boundary (region I) propagating in the $k_{1}$ mode. The electron will be (partly) transmitted to the different modes at the left side of the boundary (region II) and (partly) reflected to the left propagating modes in region I. The boundary conditions at $\pm \infty $ then yield expressions for the vectors $\mathcal{C}_{I}$ and $\mathcal{C}_{II}$ of the system in region I and II, respectively, as
\begin{eqnarray}
\mathcal{C}_{I} &=&\left[ 1,r_{1}^{1},0,r_{2}^{1},0,r_{3}^{1}\right] ^{T}, \\
\mathcal{C}_{II} &=&\left[ t_{1}^{1},0,t_{2}^{1},0,t_{3}^{1},0\right] ^{T}.
\end{eqnarray}%
In these expressions the coefficients '$r$' indicate the left propagating reflected waves and '$t$' indicates the right propagating transmitted waves. The subscripts denote the channel in which the waves are propagating and the superscripts indicate the incident mode of the wave. In Fig. \ref{Fig:GenericTransAndRefl} different possible channels are shown schematically for a generic two band system. For TLG a third band needs to be included, leading to extra combinations of scattered transmission and reflection amplitudes. Using the expression for the current density in the $x$ direction given by Eq. $\left( \ref{CurrentDensityFormula}\right) $, conservation of probability current leads to the normalization condition for the different channels as
\begin{eqnarray}
&&\mathcal{A}_{1,1}^{x}-\left\vert r_{1}^{1}\right\vert ^{2}\mathcal{A}_{1,1}^{x}-\left\vert r_{2}^{1}\right\vert ^{2}\mathcal{A}_{2,2}^{x}-\left\vert r_{3}^{1}\right\vert ^{2}\mathcal{A}_{3,3}^{x} \\
&=&\left\vert t_{1}^{1}\right\vert ^{2}\mathcal{A}_{1,1}^{x}+\left\vert t_{2}^{1}\right\vert ^{2}\mathcal{A}_{2,2}^{x}+\left\vert t_{3}^{1}\right\vert ^{2}\mathcal{A}_{3,3}^{x},
\end{eqnarray}
which allows to define the scattered transmission and reflection probabilities as
\begin{equation}
T_{j}^{i}=\left\vert t_{j}^{i}\right\vert ^{2}\frac{\mathcal{A}_{j,j}^{x}}{\mathcal{A}_{i,i}^{x}}\text{ and }R_{j}^{i}=\left\vert r_{j}^{i}\right\vert^{2}\frac{\mathcal{A}_{j,j}^{x}}{\mathcal{A}_{i,i}^{x}}.
\label{TransAndRefl}
\end{equation}
The transmission and reflection probabilities therefore depend on the value of the coefficients of the vector $\mathcal{C}$. These coefficients can be found by matching the plane wave solutions of both regions at the boundary. Using the transfermatrix approach as explained by Barbier \textit{et al.} \cite{Barbier2010}, one can model a sequence of different regions to create more complex structures such as a $pnp$ junction.

To find empirically relevant quantities, the transmission probabilities can be used to calculate the conductance which is defined by the Landauer-B\"{u}ttiker formula\cite{Blanter2000}:
\begin{equation}
G\left( E\right) =G_{0}\frac{L_{y}}{2\pi }\int_{-\infty }^{\infty
}dk_{y}\sum_{l,m=1}^{3}T_{m}^{l}\left( E,k_{y}\right) ,
\end{equation}%
with $G_{0}=4e^{2}/h$, four times the quantum of conductance due to valley and spin degeneracy and $L_{y}$ is the length of the sample in the $y-$direction.

Note that if both Dirac points are equivalent, the system is time reversal invariant. Therefore electrons that scatter reflectively from the $i^{th}$ to the $j^{th}$ band are equivalent to electrons reflecting from the $j^{th}$ in the $i^{th}$ band near the other Dirac point. A similar symmetry arises for the transmission when the potential, i.e. the eigenfunctions, of the first and last region of a series of boundaries are the same, such as with a $pnp$ junction. Transmission from the $i^{th}$ into the $j^{th}$ band near the first Dirac point is then equivalent to transmission scattering from the $j^{th}$ into the $i^{th}$ band near the other Dirac point while being incident on the opposite side of the $pnp$ junction. These equivalences, together with the equivalence between the Dirac points lead to the following symmetry in the reflection and transmission probability
\begin{equation}
R_{j}^{i}=R_{i}^{j}\text{ and }T_{j}^{i}=T_{i}^{j}\text{,}
\label{Eq:TAndREqui}
\end{equation}%
where the second equation only holds if the first and the last region are equivalent. This decreases the number of different probabilities to only 6 reflection and 9 transmission probabilities for a general potential and to 6 transmission and 6 reflection probabilities for a $pnp$ junction.

A final remark considering the symmetry of the system with respect to the sign of the angle of incidence has to be made. The Hamiltonians given in Eqs. (\ref{Eq:HamABC}) and (\ref{Eq:HamABA}) are not symmetric under the change of the sign of $k_y$. Although the electronic spectrum of these systems do have this symmetry, the obtained transmission and reflection results do not have to bare the reflection symmetry with respect to normal incidence. ABC TLG however has an additional symmetry, namely that by turning the system upside down and rotating in plane by an angle of $\pi$, the system is the same, but the sublattices $\alpha_1$ and $\beta_3$, $\beta_1$ and $\alpha_3$ and $\alpha_2$ and $\beta_2$ are interchanged. This transformation should not change the result, but it transforms the ABC Hamiltonian in such a way that $k_y \rightarrow -k_y$ implying that the obtained results should also have this symmetry. A similar argument also holds for MLG, where both sublattices are interchanged, and for BLG, where the exchange is between the sublattices connected by the interlayer hopping and those that are not connected. Therefore, the symmetry also holds for unbiased ABA TLG since its Hamiltonian consists of MLG and BLG like blocks as shown in Eq. (\ref{HamABATransfro}). These symmetry arguments are not valid anymore if a DG is applied since it breaks the interlayer sublattice equivalence of ABC TLG and the separable behaviour of ABA TLG electrons into MLG and BLG like ones. Non symmetric results are thus expected for such systems. Although being counterintuitive, the occurrence of this asymmetry is not an unphysical result since the effect is exactly opposite for electrons near the other Dirac point. In the second valley, the electrons are described by a Hamiltonian that is similar to that near the first Dirac point, but with the exchange of $E \rightarrow -E$ and $k_y \rightarrow -k_y$. The results for electrons (holes) in the first valley are therefore the same as for holes (electrons) with opposite transverse momentum in the second valley. The electron-hole symmetry of the system finally completes the argument.

\section{Numerical results}\label{Sec:Results}
In the following sections we present results obtained for both SG and DG applied to trilayer graphene samples with both ABC and ABA stacking configurations for both $pn$ and $pnp$ junctions.

\subsection{Single gated device}

\begin{figure}[tb]
\centering
\includegraphics[width = 8cm]{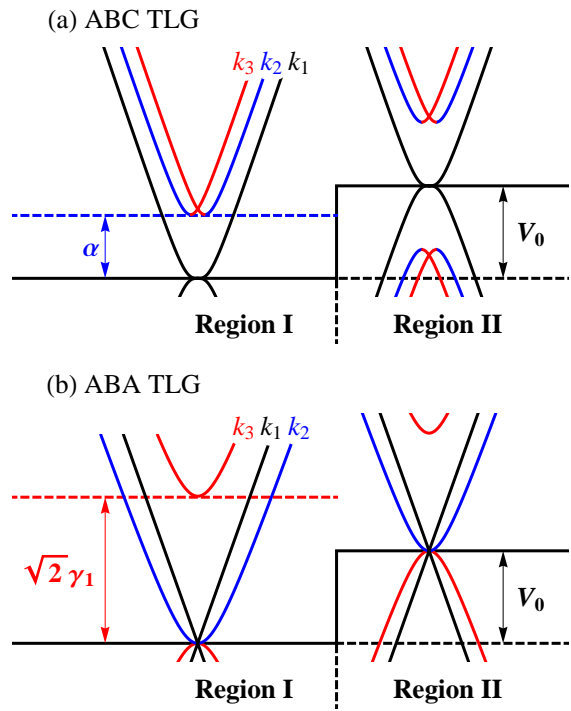}
\caption{(Colour online) Schematic representation of the energy spectrum of (a) ABC TLG and (b) ABA TLG at both sides of a SG potential boundary of height $V_0$ without interlayer bias. The colour of the curves indicate the wave vector associated with it. The minimum of the upper two bands of ABC TLG is located at energy $\alpha\approx 346meV$.}
\label{Fig:SG}
\end{figure}

ABC TLG has a maximum of three distinct modes of propagation, two of which are only propagating if the Fermi energy is large enough, i.e. $E>\alpha = \frac{3}{4}\sqrt{\frac{3}{2}}\gamma_{1}\approx 346meV$. In Fig. \ref{Fig:SG}(a) the energy spectrum of ABC TLG is shown at both sides of the boundary. The branches of the spectrum are coloured corresponding to the modes of propagation.

\begin{figure}[tb]
\centering
\includegraphics[width = 8cm]{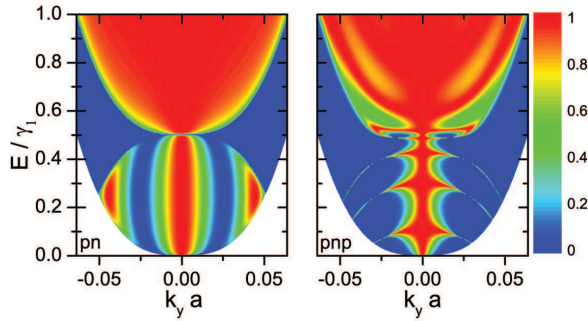}
\caption{(Colour online) Transmission probability as function of the energy and transverse wave vector through (left) a SG $pn$ junction of height $V=0.5\protect\gamma_1$ and (right) a SG $pnp$ junction of width $d=25nm$ and the same height using the two band Hamiltonian.}
\label{Fig:RedTrans}
\end{figure}

At low energy, there is only one mode of propagation in ABC TLG which allowed before to introduce the two band Hamiltonian, Eq. $\left( \ref{TwoBandTLG}\right) $, that approximates the system as one with a cubic energy momentum relation. In Fig. \ref{Fig:RedTrans} we show the transmission probability through a SG $pn$ and a $pnp$ junction for low energy in case of the two band system. Notice that at normal incidence, i.e. $k_{y}=0$, the transmission equals unity independent of energy or width of the barrier. This Klein tunneling is the consequence of conservation of pseudospin and occurs for rhombohedrally stacked multilayers with an odd number of layers. Furthermore, there is a region of $k_{y}$ corresponding to an angle of incidence of $\pm \phi=\pi/6$, for which the transmission is suppressed when $E<V_0$. This is another consequence of the pseudospinorial nature of the electrons in ABC TLG. At this angle of incidence, the propagating states inside the junction are disconnected from those outside resulting in a lower transmission.

At non normal incidence for $E<V$, very narrow resonances show up for the $pnp$ junction (see Fig. \ref{Fig:RedTrans}) that are similar to the Fabry-P\'{e}rot resonances observed in MLG\cite{RamezaniMasir2010} and BLG\cite{VanDuppen}. The number of resonances and their energy and $k_{y}$ dependence vary with the width of the barrier. For the $pn$ junction, spots of high transmission are found at non normal incidence. These spots are also a consequence of the chiral nature of the charge carriers and occur near an incident angle of $\phi =\pi /3$.\cite{VanDuppen2013}

When the Fermi energy $E$ of the electrons under consideration is larger than $\alpha $, i.e. the minimum of the second band in Fig. \ref{Fig:SG}(a), the second and third modes of propagation become accessible. For these energies, the two band approximation is not sufficient anymore and interband scattering between the three bands needs to be taken into account.

\begin{figure*}[tbh]
\begin{center}
\includegraphics[width=16cm]{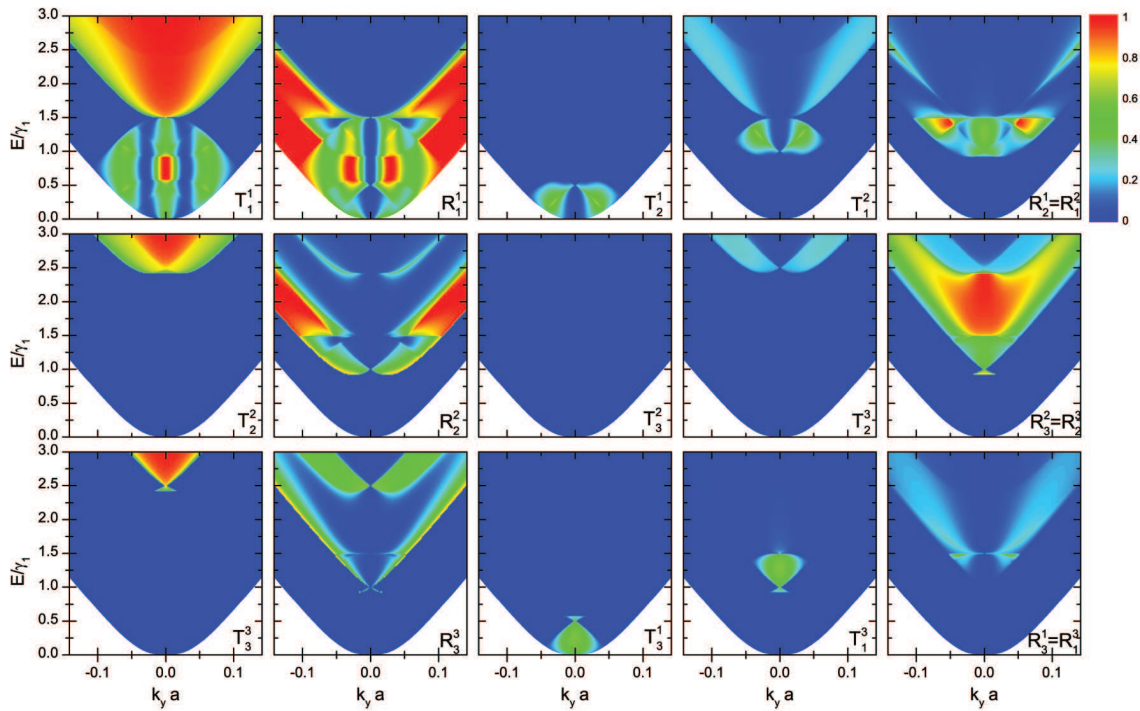}
\end{center}
\caption{(Colour online) Transmission and reflection probabilities for a single gated $pn$ junction on ABC TLG of height $V_{0}=1.5 \protect\gamma _{1}$ as function of the energy and transverse momentum $k_y$. }
\label{StepABC}
\end{figure*}

In Fig. \ref{StepABC} we show the transmission and reflection channels for a $pn$ junction of height $V_{0}=1.5\gamma _{1}$ as function of the transverse wave vector and the Fermi energy of the incident electron using the six band Hamiltonian. Note that Klein tunneling and cloaking when described by the two band approximation reoccur in the energy interval where only one band is propagating in both regions. This can be seen in the $T_{1}^{1}$ and $R_{1}^{1}$ channel in the energy interval $V_{0}-\alpha <E<\alpha $. Outside this range, for lower energy the $k_{2}$ and $k_{3}$ states inside the junction are propagating, giving rise to non zero scattered transmission probabilities $T_{2}^{1}$ and $T_{3}^{1}$ which reduces the transmission via the $k_{1}$ channel. For larger energy, it is possible to reflect in the left propagating $k_{2}$ and $k_{3}$ states in region I which further reduces the direct transmission. For Fermi energy larger than the junction's height, the $T_{1}^{1}$ channel coincides with the two band calculation. Notice however that electrons impinging in the second or third band, are reflectively scattered in the third and second band respectively, rather than propagating in the $k_{1}$ channel.

\begin{figure*}[tbh]
\begin{center}
\includegraphics[width=16cm]{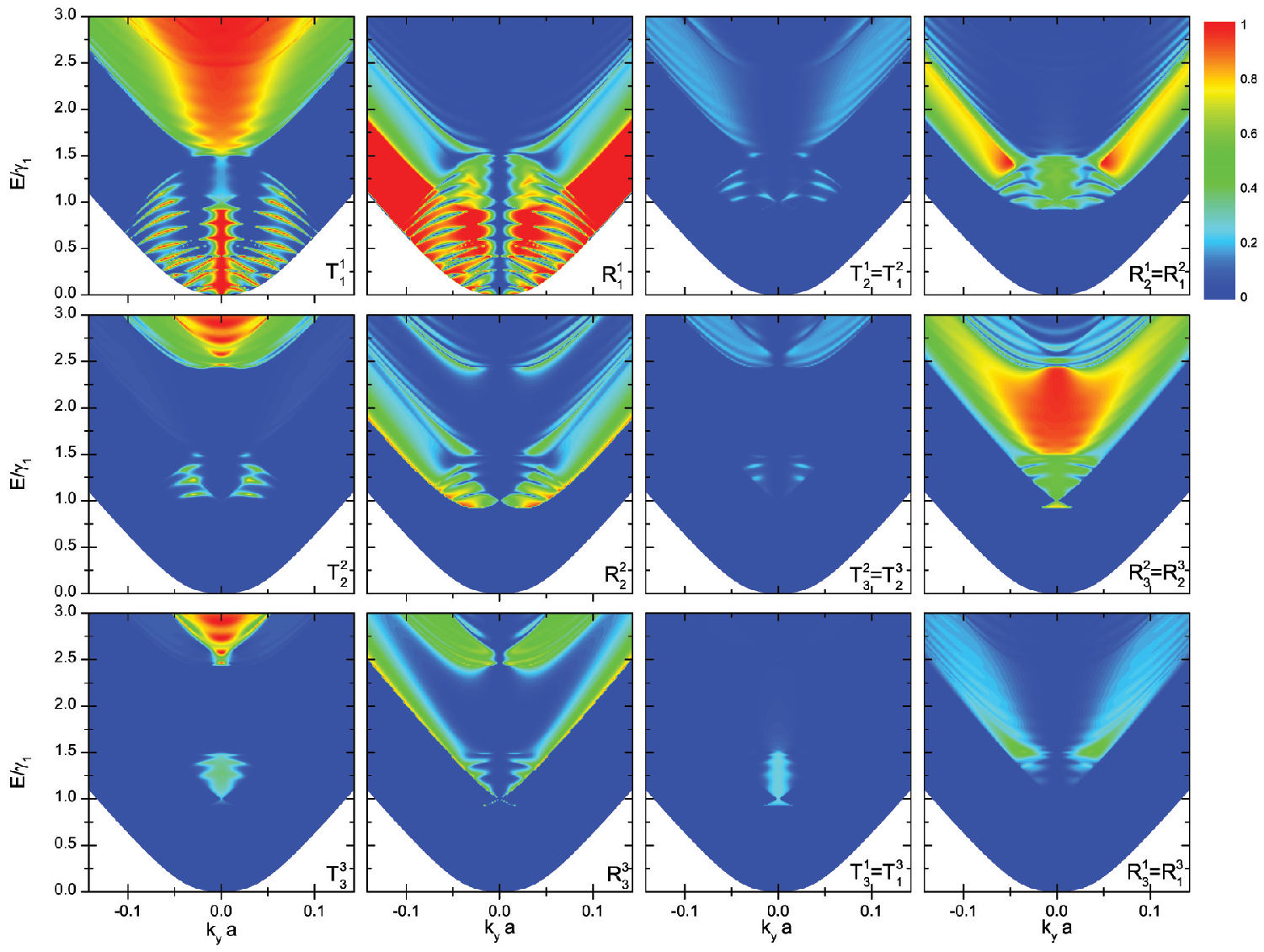}
\end{center}
\caption{(Colour online) Transmission and reflection probabilities for a single gated $pnp$ junction on ABC TLG of height $V_{0}=1.5 \protect\gamma _{1}$ and width $d=25nm$ as function of the energy and transverse momentum $k_y$. }
\label{25nmHighV}
\end{figure*}

In Fig. \ref{25nmHighV} the twelve transmission and reflection probabilities for electrons incident on a SG $pnp$ junction of width $d=25nm$ and height $V=1.5\gamma _{1}$ using the six band Hamiltonian are shown. The results for the $T_{1}^{1}$ and $R_{1}^{1}$ channels are similar to those obtained from the two band Hamiltonian. Notice again Klein tunneling at normal incidence and the suppression due to cloaking at non normal incidence. Outside the energy range of validity for the two band approximation, i.e. $E<V-\alpha$ and $E>\alpha$, the structure of the resonances at low energy is a superposition of the one similar to the two band system and another type of resonances. The latter are resonances due to propagation via the $k_{2}$ and $k_{3}$ bands inside the junction region and therefore only show up when $E<V_0-\alpha$. Furthermore, reflective scattering is large when all three modes are evanescent in the junction region. This is analogous to the interband scattering for BLG.\cite{VanDuppen} Another striking fact is that at normal incidence when a second mode of propagation is possible, Klein tunneling is suppressed in favor of scattered reflection in the $R_{2}^{1}$ and $R_{3}^{1}$ channels. This leads to the argument that Klein tunneling can be seen as the consequence of a suppression of the ability to backscatter rather than the link between forward propagating inside and outside the potential barrier.

The large scattered reflection between the $k_{2}$ and $k_{3}$ channels shown in Fig. \ref{25nmHighV} at normal incidence is a consequence of the tight relation between the $k_{2}$ and $k_{3}$ propagation modes. When Eq. $\left( \ref{DetEq}\right) $ is solved for normal incidence, one finds an expression consisting of three hyperbolic bands, two of which intersect at $E=\gamma _{1}$. At normal incidence, the reflected branch of the $k_{3}$ spectrum as indicated in Fig. \ref{Fig:SG}(a) therefore corresponds with the forward propagating $k_{2}$ band and vice versa. Although at non normal incidence this relation is no longer valid, the reflective scattering remains strong at near normal incidence.

\begin{figure}[tb]
\centering
\includegraphics[width = 8cm]{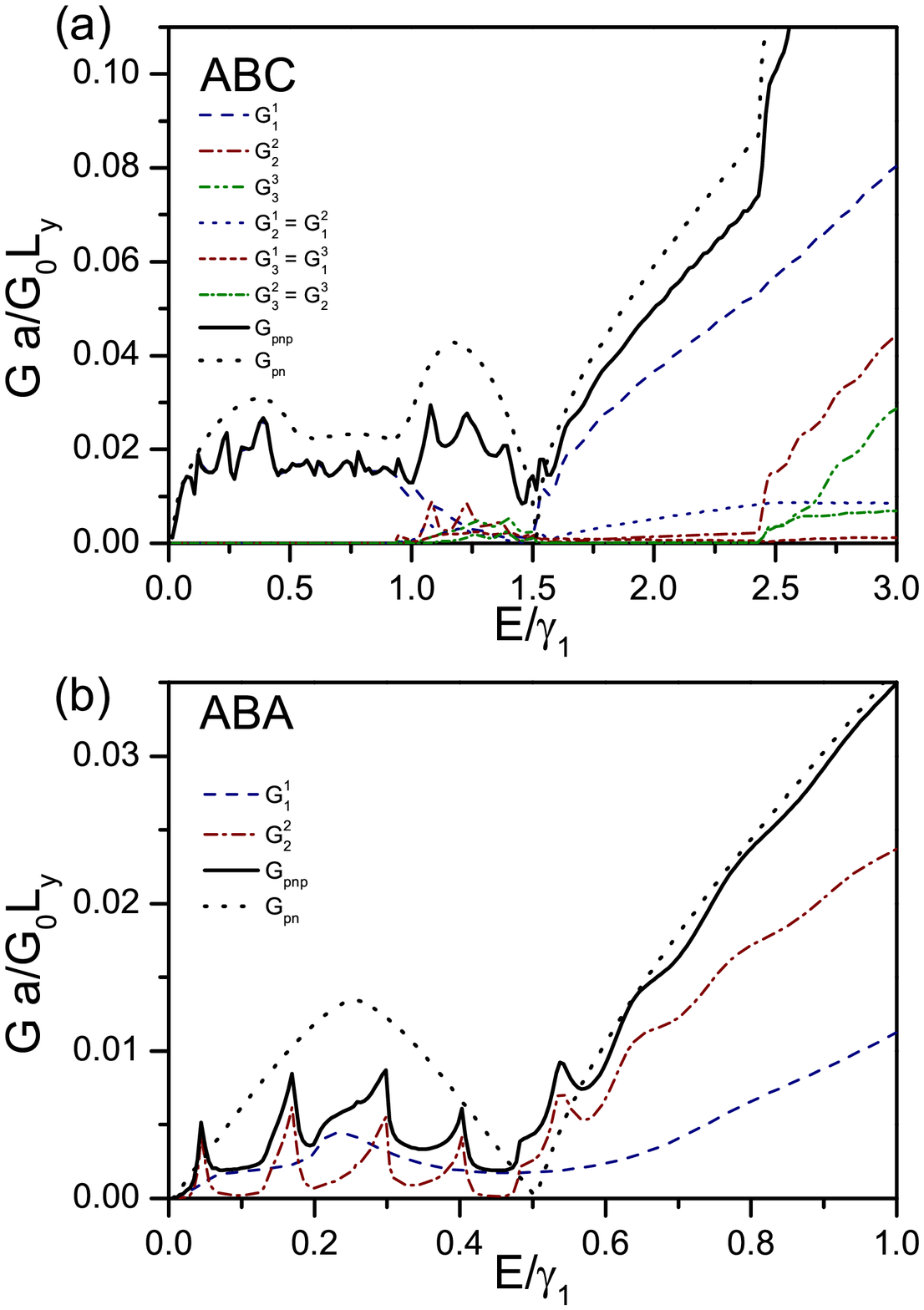}
\caption{(Colour online) Energy dependency of the conductance at a SG junction for (a) ABC TLG with potential height $V_0=1.5\gamma_1$ and (b) ABA TLG with potential height $V_0=0.5\gamma_1$. The solid black curves correspond to the total conductance through a $pnp$ junction of length $d=25nm$, the dashed coloured curves indicate the contributions of the different propagation channels. The dotted black curve corresponds to the conductance through a $pn$ junction of the same height.}
\label{Fig:ConductanceSG}
\end{figure}

In Fig. \ref{Fig:ConductanceSG}(a) we show the energy dependence of the conductance for the same system as for the previous results. The resonances that are visible in the transmission probability show up as peaks in the conductance. Furthermore, the results indicate the availability of additional propagation modes via the $k_{2}$ and $k_{3}$ channels for higher energy resulting in a clear signature of higher conductance. The increased conductance for the high barrier is a distinct feature from the results of the two band approximation. For low energy ($E<0.5\gamma _{1}$) and just below the barrier's height ($\gamma _{1}<E<V$), the conductance is raised due to propagation via the second and third channels in the barrier region. This is absent in the two band approximation.

The SG imposes a potential that treats each layer in the same way and thus keeps the existing symmetries of the system. As a consequence, the spectrum of the system is only shifted by an overall potential term as shown in Fig. \ref{Fig:SG}(b). The monolayer-like and bilayer-like propagation in ABA TLG therefore remains the same and so the electrons are described as if they propagate in MLG or BLG. The transmission and reflection through $pn$ and $pnp$ SG devices on MLG and BLG has already been investigated in depth\cite{Katsnelson2006, RamezaniMasir2010, VanDuppen}. It was pointed out that the MLG electrons exhibit Klein tunneling when they hit the boundary perpendicularly. They are transmitted with unit probability, irrespective of the height $V_0$ or the width $d$ of the gated region. However, BLG electrons are cloaked from the propagating states at this angle of incidence and therefore their transmission is suppressed\cite{Gu2011}. At non normal incidence, cloaking and Klein tunneling can occur if certain conditions are satisfies as described recently\cite{VanDuppen2013}. The MLG and BLG like propagation channels have each their own separate contribution to the conductance of an ABA TLG sample. In Fig. \ref{Fig:ConductanceSG}(b) we show the energy dependency of the conductance of an ABA TLG sample with a SG $pn$ and $pnp$ junction. From the contributions of the separate channels one finds that when $E<V_0$, the MLG conductance $G_1^1$ is larger than that of the BLG due to the Klein tunneling, but the BLG conductance $G_2^2$, although being cloaked, shows clear peaks reminiscent from the BLG transmission resonances. For larger energy, when $E>V_0$, the BLG contribution is however larger than that of the MLG. The overall conductance of a $pn$ junction doesn't differ much from the $pnp$ junction when $E>V_0$, while it misses the resonances at energy lower than the junction's height.

\subsection{Double gated device}

\begin{figure}[tb]
\centering
\includegraphics[width = 8cm]{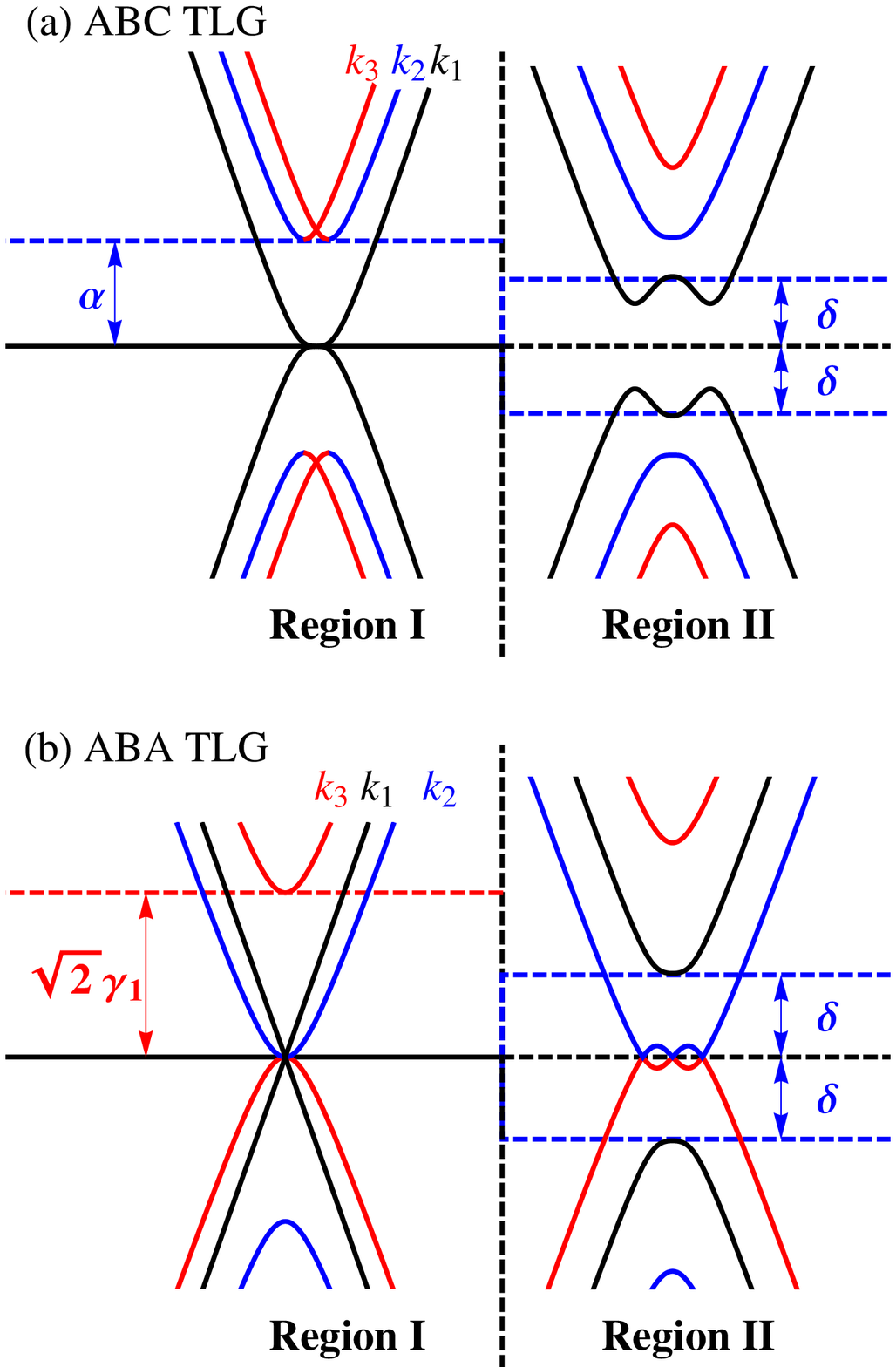}
\caption{(Colour online) Schematic representation of the energy spectrum of (a) ABC TLG and (b) ABA TLG at both sides of a DG boundary with interlayer bias $\delta$. The colour of the curves indicate the wave vector associated with it.}
\label{Fig:DG}
\end{figure}

\begin{figure}[tb]
\centering
\includegraphics[width = 8cm]{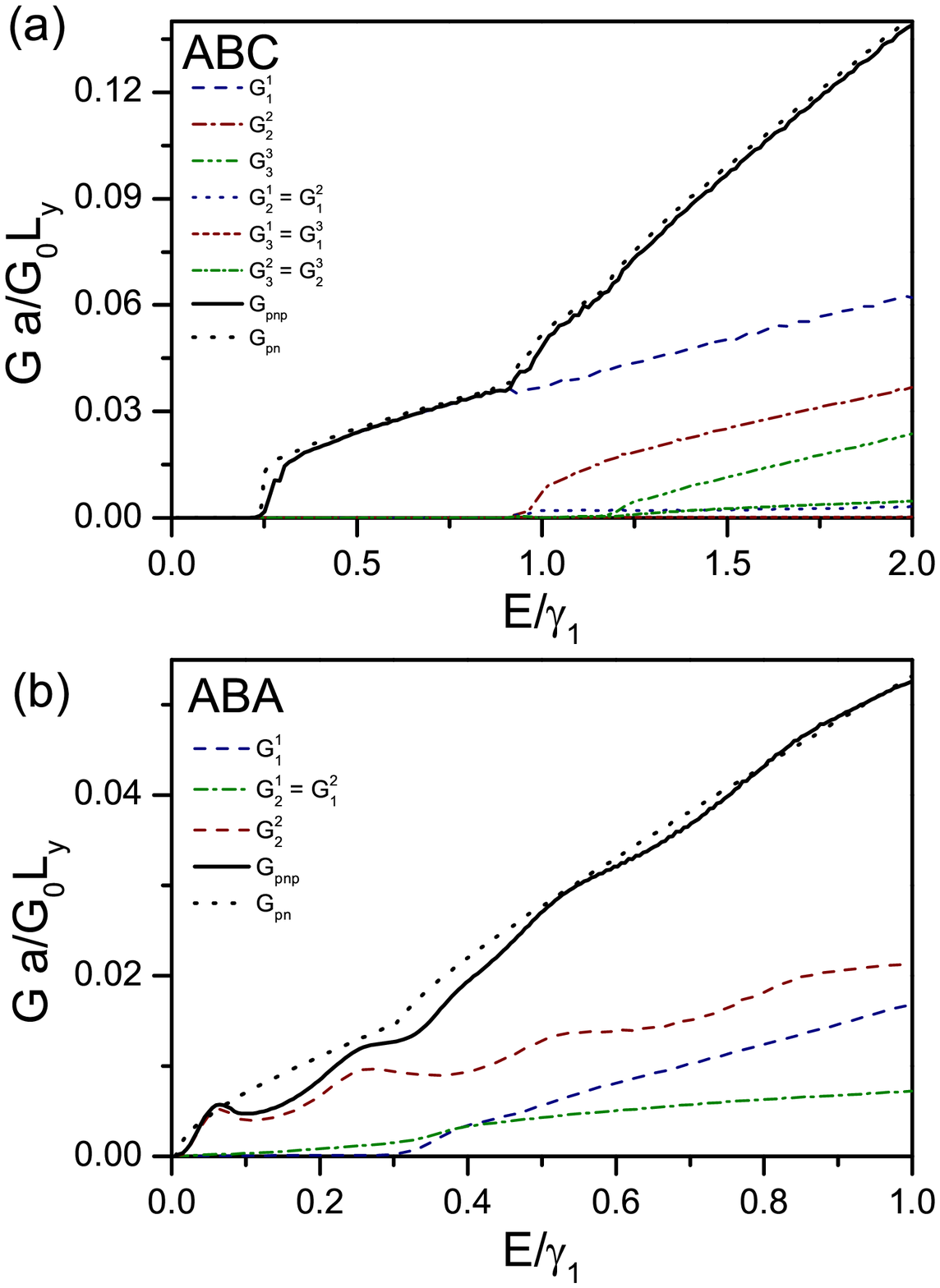}
\caption{(Colour online)  Energy dependence of the conductance at a DG junction of strength $\delta=0.3 \gamma_1$ for (a) ABC TLG and (b) ABA TLG. The solid black curves correspond to a $pnp$ junction of length $d=25nm$, the dashed coloured curves indicate the contributions of the different propagation channels. The dotted black curve corresponds to the conductance through a $pn$ junction of the same strength. }
\label{Fig:ConductanceWrong}
\end{figure}

A potential difference between the layers of an ABC TLG destroys the degeneracy of the valence and conduction band at the Dirac point by opening up a band gap between them\cite{Avetisyan2010, Craciun2009}. This is illustrated in Fig. \ref{Fig:DG}(a). Due to the absence of propagating states inside the band gap, the transmission is reduced in all channels which can be seen by the low conductance as shown in Fig. \ref{Fig:ConductanceWrong}(a) where both the $pn$ DG and $pnp$ DG conductance is shown. When the propagating states become available, the conductance increases sharply to unity in the regime where only one band is available to conduct. As more bands become possible to conduct, additional transmission channels contribute to the overall conductance increasing it to almost perfect conductance. Due to the absence of resonances, the $pn$ DG and $pnp$ DG conductance are similar, only showing the effect of additional transmission channels. The DG lifts the valley degeneracy and therefore the symmetry in the scattered transmission expressed in Eq. (\ref{Eq:TAndREqui}) is no longer valid but are the same up to a reflection with respect to normal incidence. The conductance however sums over both positive and negative angles so the effect is absent. Since the asymmetric effect is only possible when several modes of propagation are available, it will be more pronounced when considering ABA TLG.

The breaking of the interlayer symmetry induced by a DG couples the MLG and BLG modes as shown in Eq. (\ref{HamABATransfro}). We have calculated the scattering between the linear and the lowest hyperbolic band due to a $pn$ DG and a $pnp$ DG. The way in which the ABA spectrum is influenced by the DG is shown schematically in Fig. \ref{Fig:DG}(b). A band gap is created between the linear bands for $E<\delta$, the hyperbolic bands however remain gapless touching linearly at the Dirac point and intersecting in a circle of radius $k=\delta/\gamma_1 a$ around the Dirac point. For low energy, this feature allows for three conduction modes of the same band while for a Fermi energy that is a little higher, only one mode is available.

\begin{figure*}[tb]
\centering
\includegraphics[width = 18cm]{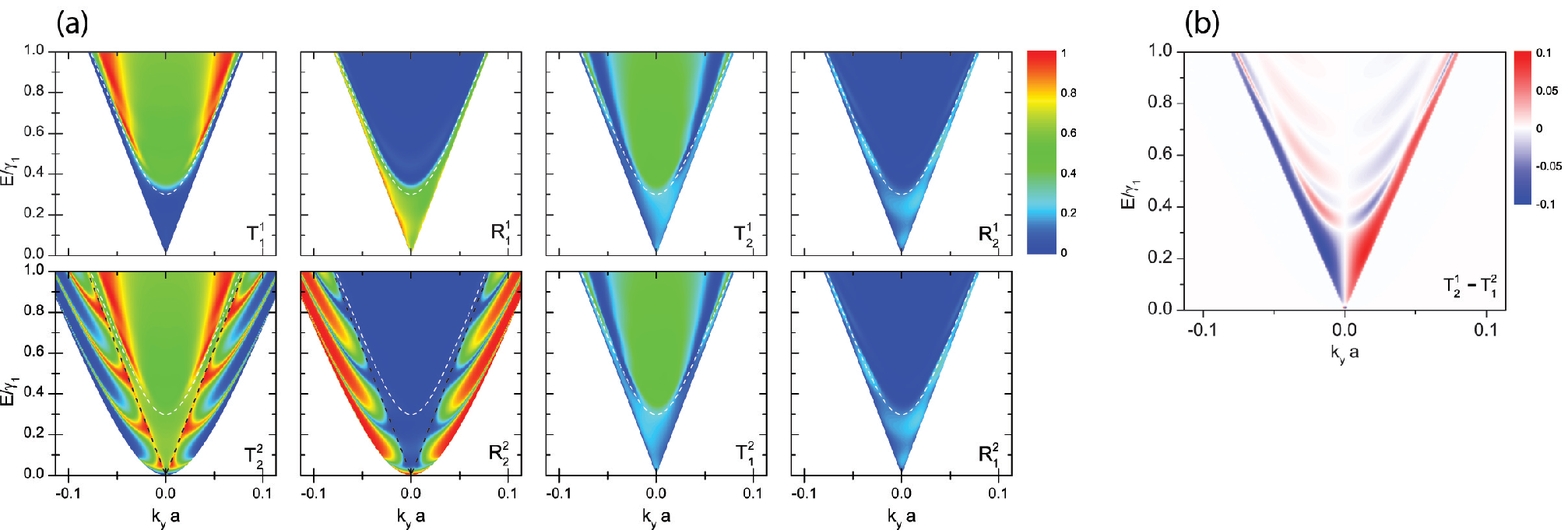}
\caption{(Colour online) (a) Contourplots of the transmission and reflection probability for a DG $pnp$ junction in ABA TLG of width $d=25nm$ and strength $\delta=0.3\gamma_1$ as function of the transverse wave vector and the Fermi energy of the incident electron. The dashed lines indicate the boundaries of the regions were different modes are available to propagate. (b) Difference between the scattered transmissions $T_2^1$ and $T_1^2$.}
\label{Fig:ABATransGrid}
\end{figure*}

In Fig. \ref{Fig:ABATransGrid}(a) we show the angular and energy dependency of the transmission and reflection probabilities at a $pnp$ DG. They can be divided in several regions defined by which mode is propagating as shown by the dashed lines superimposed on the results. A clear feature is the hyperbola reminiscent of the gapped Dirac cone due to the interlayer asymmetry in the junction region which is displayed as a white dashed curve. Inside this hyperbola, it is possible to propagate through the $k_1$ channel and this increases the linear to linear energy band $T_1^1$ transmission. In contrast to MLG electrons, it is possible to be backscattered in the linear band. A second clear feature is the Dirac cone defining the MLG propagating states outside the junction, which is indicated by a black dashed curve. Outside the Dirac cone, only the BLG-like states are propagating both inside and outside the junction. This leads to angular depending resonances that vary with the length of the $pn$ junction. The previously discussed angular asymmetry is clearly seen in the reflection channels and the scattered transmission channels. While the scattered reflection keeps the interband symmetry $R_2^1=R^2_1$, the scattered transmission differs. In Fig. \ref{Fig:ABATransGrid}(b) we show the difference between both scattered reflection channels. The result makes clear that the scattered transmission is the same under a flip of the sign of $k_y$, i.e.
$T^1_2(k_y)=T^2_1(-k_y)$. This is an immediate consequence of the time reversal symmetry of the system. As mentioned above, an electron scattering from band 1 to band 2, near the Dirac point $K$, $T_{2,K}^1$, is equivalent to one scattering from band 2 to band 1 near the other Dirac point $K'$, $T_{1,K'}^2$. Since the latter is the same as $T_{1,K}^2$ with a sign flip in $k_y$, the scattered transmissions near the same Dirac point are the same under a change of sign in $k_y$. Note that these asymmetric results are not carried through in the conductance.

\begin{figure}[tb]
\centering
\includegraphics[width = 8cm]{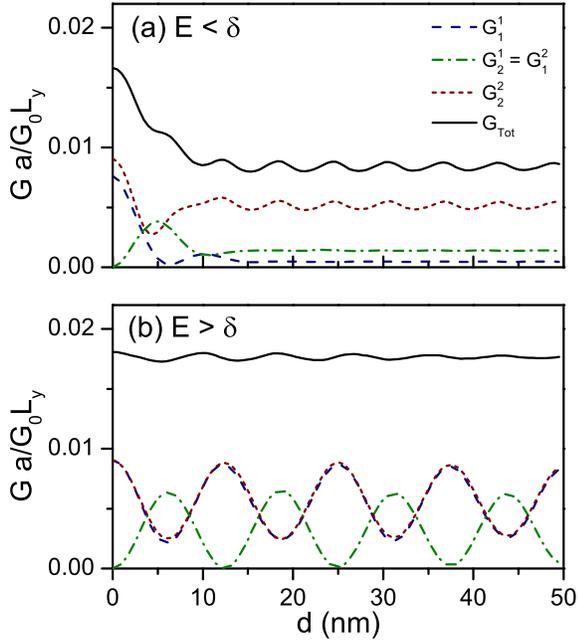}
\caption{(Colour online) Conductance as function of the width of the $pnp$ DG of strength $\delta=0.5\gamma_1$ for Fermi energy (a) $E=0.3 \gamma_1$ and (b) $E=0.8\gamma_1$.}
\label{Fig:TransLength}
\end{figure}

In Fig. \ref{Fig:ConductanceWrong}(b) we show the conductance for a $pn$ and a finite width $pnp$ DG in ABA TLG. The resonances caused by the $T_2^2$ channels show up as small bumps in the conductance when the Fermi energy is below $\delta$. For higher energy, the contribution of the MLG like mode of propagation is marked by a rise in conductance due to the contribution of the $G_1^1$. In Fig. \ref{Fig:TransLength} the conductance is shown as function of the width of the $pnp$ DG with strength $\delta=0.5\gamma_1$ for different levels of the Fermi energy. In Fig. \ref{Fig:TransLength}(a), we show the conductance for a Fermi energy $E<\delta$. Now there is only one mode of propagation in the junction and the oscillatory behaviour is only due to the resonances of the $G_2^2$ term. The direct $G_1^1$ contribution diminishes exponentially since and obtains a finite value independent of the width of the junction, similar to the interband scattering contributions. Fig. \ref{Fig:TransLength}(b) shows the conductance for $E>\delta$. Now the contributions of both direct channels, $G_1^1$ and $G_2^2$, are almost equal and oscillate with the width of the junction while the scattered transmission oscillates oppositely. In this way, one can determine the amount of electrons scattered between the monolayer- and bilayer-like bands by applying a DG of the correct strength and width.

\section{Conclusion}\label{Sec:Conclusion}
In this paper we have studied the electronic transport in trilayer graphene through $pn$ and $pnp$ junctions consisting of a single and a double gate. For ABC TLG we have shown that the availability of different modes of propagation at higher energy leads to typical features in the conductance and that Klein tunneling occurs only if there is no other propagation mode available. Furthermore, we have calculated the effect of the band gap induced by a DG on ABC TLG in the conductance and shown that the conductance is nearly zero in the gapped region.

We also modeled the effect of SG and DG gates on ABA TLG and found that the SG behaviour is a superposition of a monolayer and bilayer like system. The DG however mixes both types of bands and breaks the angular symmetry with respect to normal incidence. This peculiar result emphasis the necessity to include both Dirac points even if intervalley scattering is prohibited since the electron behaviour near the other Dirac point restores the symmetry. The DG finally allows electrons to scatter between the monolayer and bilayer like bands and we have shown that the strength of scattering depends on the with the width of the $pnp$ junction and the Fermi energy considered.

\section{Acknowledgments}

This work was supported by the European Science Foundation (ESF) under the EUROCORES Program Euro-GRAPHENE within the project CONGRAN, the Flemish Science Foundation (FWO-Vl) and the Methusalem Foundation of the Flemish Government.

\end{document}